\documentclass[a4paper]{jpconf}
\usepackage{graphicx}\usepackage{amssymb}
\newcommand{\be}{\begin{equation}} 
\newcommand{\ee}{\end{equation}}
\newcommand{\bc}{\begin{center}}
\newcommand{\ec}{\end{center}}
\begin{document}

\title{Applications of nuclear physics to a wider context:
from molecules to stars passing through hypernuclei
}
\author{Lorenzo Fortunato}
\address{Dipartimento di Fisica e Astronomia ``G.Galilei'', via Marzolo 8, I-35131 Padova, Italy}
\address{I.N.F.N., Sezione di Padova, via Marzolo 8, I-35131 Padova, Italy}
\ead{fortunat@pd.infn.it}

\begin{abstract}
In this contribution I will review some of the researches that are currently 
being pursued in Padova (mainly within the In:Theory and Strength projects), focusing on the interdisciplinary applications
of nuclear theory to several other branches of physics, with the aim 
of contributing to show the centrality of nuclear theory in the Italian scientific 
scenario and the prominence of this fertile field in fostering new physics.

In particular, I will talk about: i) the recent solution of the long-standing "electron screening puzzle" that 
settles a fundamental controversy in nuclear astrophysics between the 
outcome of lab experiments on earth and nuclear reactions happening in stars;
 the application of algebraic methods to very diverse systems such as: ii) the 
supramolecular complex H2@C60, i.e. a diatomic hydrogen molecule caged in a
fullerene and iii) to the spectrum of hypernuclei, i.e. systems made of a Lambda 
particles trapped in (heavy) nuclei.
\end{abstract}

\section{Introduction}
The aim of the present manuscript is to summarize part of the research lines that have been recently pursued in Padova. While most of these could be classified as nuclear theory (Strength project of the INFN section in Padova), interdisciplinary research has been pursued in a broader field that goes from nuclear astrophysics to molecular and to subnuclear physics.
We will give three examples, one related to the solution of the famous electron screening puzzle in nuclear astrophysics\cite{spit,news}, one related to application of algebraic theories to diatomic molecules caged in a fullerene \cite{h2} and one to an algebraic theory for $\Lambda$ particles in nuclei \cite{hyp}.

\section{Electron screening puzzle}
Nuclear reactions in stars happen at energies around the Gamow peak  $E_G <<  E_{C.B.}$, much lower than the energy of the Coulomb barrier, at the presence of an electron plasma that creates a screening effect that enhances the reaction cross-sections at these ultra-low energies. At these energies tunneling of the potential barrier becomes dominant and the typical exponential behaviour of the cross-section with energy can be removed by defining the S-factor as:
\begin{equation}
S(E) = E \sigma (E) e^{2\pi\eta(E)}
\end{equation}
where $\eta= Z_1Z_2 e^2/\hbar v$ is the Sommerfeld parameter that depends on the charges of the two reactants and their relative velocity.
\begin{table}
\caption{Theoretical adiabatic limits, $U_e^{adlim}$, and experimental values of the electron screening potentials, $U_e^{exp}$, for a number of reactions. See Ref. \cite{spit} for more details and references.}
\begin{center}
\begin{tabular}{lllc}
\hline
           {}&                                              {}&                                  {}&                                            \\[.1ex]	
            {}&Reaction      				  {}&$U_e^{adlim}$     	   {}&$U_e^{exp}$               {}      	       \\[.1ex]  
            {}&                  					  {}& (eV)                 	   {}&(eV)                             {} \\[.25ex]           
\hline
$(1)$  {}&$^2$H($d,t$)$^1$H                                                   {}&14       {}&13.2$\pm$1.8        {}   \\[.1ex]         
$(2)$  {}&$^2$H($d$,$^3$He)n                                               {}&14       {}&11.7 $\pm$1.6       {}  \\[.1ex]	    
$(3)$ {}&$^3$He($d$,$p$)$^4$He                                         {}&65        {}&109$\pm$9           {}   \\[.1ex]   
$(4)$ {}&$^3$He($d$,$p$)$^4$He                                         {}&120      {}&219$\pm$7            {} \\[.1ex]       
$(5)$ {}&$^3$He($^3$He,2p)$^4$He                                    {}&240     {}&294$\pm$47          {}   \\[.1ex]
$(6)$ {}&$^3$He($^3$He,2p)$^4$He                                     {}&240     {}&305$\pm$90          {}   \\[.1ex]
\hline
$(7)$  {}&$^6$Li($d$,$\alpha$)$^4$He                   	         {}&175     {}&330$\pm$120         {}     \\[.1ex]
$(8)$  {}&$^6$Li($d$,$\alpha$)$^4$He               	         {}&175     {}&380$\pm$250                {} \\[.1ex]
$(9)$  {}&$^6$Li($d$,$\alpha$)$^4$He           		         {}&175     {}&320$\pm$50         {}      \\[.1ex]
$(10)$ {}&$^6$Li($p$,$\alpha$)$^3$He                 		{}&175     {}&440$\pm$150         {}   \\[.1ex]
$(11)$ {}&$^6$Li($p$,$\alpha$)$^3$He                	         {}&175     {}&470$\pm$150         {}  \\[.1ex]
$(12)$ {}&$^6$Li($p$,$\alpha$)$^3$He                   		{}&175     {}&355$\pm$100         {}       \\[.1ex]                       	
$(13)$ {}&$^7$Li($p$,$\alpha$)$^4$He              	         {}&175     {}&300$\pm$160          {}    \\[.1ex] 
$(14)$ {}&$^7$Li($p$,$\alpha$)$^4$He             	          {}&175     {}&300$\pm$280          {} \\[.1ex]   
$(15)$ {}&$^7$Li($p$,$\alpha$)$^4$He              		{}&175     {}&425$\pm$60            {}          \\[.1ex]      
\hline
$(16)$  {}&$^9$Be($p$,$\alpha_0$)$^6$Li                        {}&240       {}&900$\pm$50             {}          \\[.1ex]                                                                                   
$(17)$  {}&$^9$Be($p$,$\alpha_0$)$^6$Li                        {}&240       {}&676$\pm$86             {}   \\[.1ex]                                                                                            
$(18)$  {}&$^{10}$B($p$,$\alpha_0$)$^7$Be                    {}&340      	{}&430$\pm$80            {}   	\\[.1ex] 
$(19)$  {}&$^{10}$B($p$,$\alpha_0$)$^7$Be                    {}&340      	{}&240$\pm$200          {}      \\[.1ex] 		 
$(20)$  {}&$^{11}$B($p$,$\alpha_0$)$^8$Be   	             {}&340       {}&430$\pm$80            {}   \\[.1ex]  
$(21)$  {}&$^{11}$B($p$,$\alpha_0$)$^8$Be       	        {}&340       {}&472$\pm$120          {}      \\[.2ex] 
\hline
\end{tabular}
\end{center}
\end{table}
Target materials in laboratory also have electrons bound in atoms, molecules or crystals' bands. Therefore the laboratory cross-sections, $\sigma_s$, and that in the stellar environment usually differ with each other and they are enhanced with respect to the cross-sections obtained from bare nuclei $\sigma_b$ (these could, in principle, be measured in beam-on-beam experiments). It is mandatory that the bare values should be extracted from beam-on-target experiments, in order to understand the effects of screening and to be able to relate the laboratory and the stellar cross-sections. An enhancement factor $f_{lab}(E)$ is usually defined \cite{Rolfs}:
\begin{eqnarray}
f_{lab}(E) = \frac{ \sigma_s(E)}{\sigma_b(E)} = \frac{S_s(E)}{S_b(E)}  \sim \exp\left[\pi\eta\frac{U_e{^{(lab)}}}{E}\right],  \label{flab}
\end{eqnarray}
that depends on the electron screening potential in laboratory experiments. The idea is that one measures the laboratory value, $\sigma_s$, obtains the bare cross-section and from that guesses the stellar value by applying a plasma enhancement factor, $\sigma_{pl}(E)=\sigma_b(E) f_{pl}$, according to the Debye-H\"uckel theory, that depends on temperature, density and other variables of the stellar environment. 
Often it is impossible to account for the large values measured in direct experiments with available atomic physics models and this discrepancy has become known as the 'electron screening puzzle'. The resolution of this conundrum is very important for applications of nuclear reactions to astrophysics and deserves special attention. In Ref. \cite{spit}, we have shown how this puzzle can be explained in terms of nuclear structure effects, namely clusterization of one or both reaction partners. The idea that nuclei exhibiting a more or less pronounced cluster structure are related to the electron screening comes from  
the observation that the list of nuclei with large discrepancies between the experimental and the theoretical (atomic physics models) values of the screening potentials contains exactly those nuclei for which a cluster structure has been proposed or identified (See Table I). If the effect is absent (or almost absent) in the first group of reactions involving Z=1 species, it becomes important for reaction involving $^{6,7}$Li and Be that are known to possess cluster structures in their ground states. This phenomenon
occurs in those nuclei where a sort of molecular bound state between two clusters is energetically favored with respect to a spherical shell-model configuration.
Not all nuclei have cluster states, nor cluster configurations, whenever they occur, necessarily take up to 100\% probability, but when they are present, they significantly alter the dynamics of fusion reactions (and others such as breakup, radiative capture processes, etc. \cite{lith, mason}).

\begin{figure}
\begin{center}
\includegraphics[scale=1.2, bb= 0 -40 175 80]{coord.pdf}~~~
\includegraphics[scale=0.4, bb= 0 10 300 600]{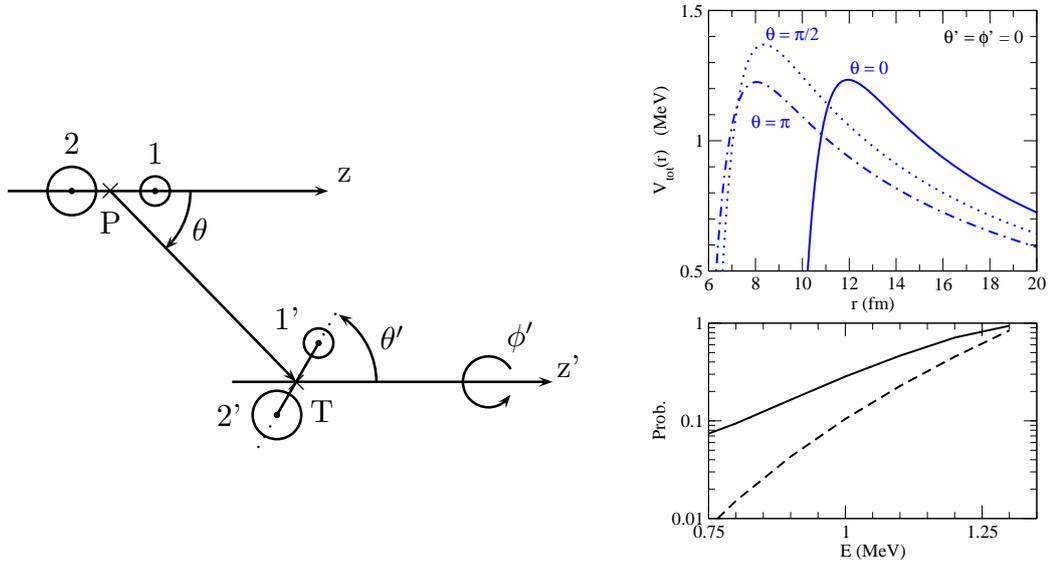}
\end{center}
\caption{Coordinate system; ion-ion potentials with three different orientations as a function of c.m. distance to show the change in barrier height and position; penetration probability of the clusterized systems (black) compared with sphere-on-sphere (dashed).\label{fign}}
\end{figure} 

The position and shape of the potential barrier thus depend on the geometry of the cluster structure (left panel of Fig. \ref{fign} for an example of parameterization of two dicluster nuclei impinging on each other), i.e. on the sum of Coulomb and nuclear potentials between each pair of fragments as
\begin{equation}
V_{tot}(r,\theta,\theta',\phi')=\sum_{i,j=1}^2 \Biggl( \frac{Z_iZ_je^2}{r_{ij}} + V_N(r_{ij})   \Biggr),
\end{equation}
where the function depends on the relative inter-cluster distance $d$, that we keep constant (and equal to $\sim 3.85$ fm obtained from the cluster model formula (A.4) of Ref. \cite{mason} ).  The tunneling probability can be calculated, at energies below the barrier, in the WKB approximation as $P=e^{-2G}$, where the Gamow tunneling factor is given by
\begin{equation}
 G(E,\theta,\theta',\phi')= \frac{\sqrt{2m}}{\hbar} \int_a^b \sqrt{V_{tot}(r,\theta,\theta',\phi') -E}~ dr \ .
\end{equation}
The angle-averaged penetration probability as a function of bombarding energy is displayed in the last panel of Fig. \ref{fign}. It is clearly larger than the analogous calculations for sphere-on-sphere and shows that the effects of clusterization on nuclear fusion at stellar energies {\it cannot be disregarded}. These effects become less pronounced at higher energies.
Further studies and experiments are needed before one can say that we have a full understanding of these phenomena, but a first significant advance toward the 
resolution of the electron screening puzzle has been achieved by recognizing the role of cluster effects.

\section{Endohedrally confined diatomic molecules}
\begin{figure}[!b]
\begin{center}
\includegraphics[scale=0.4, bb= 0 0 0 300]{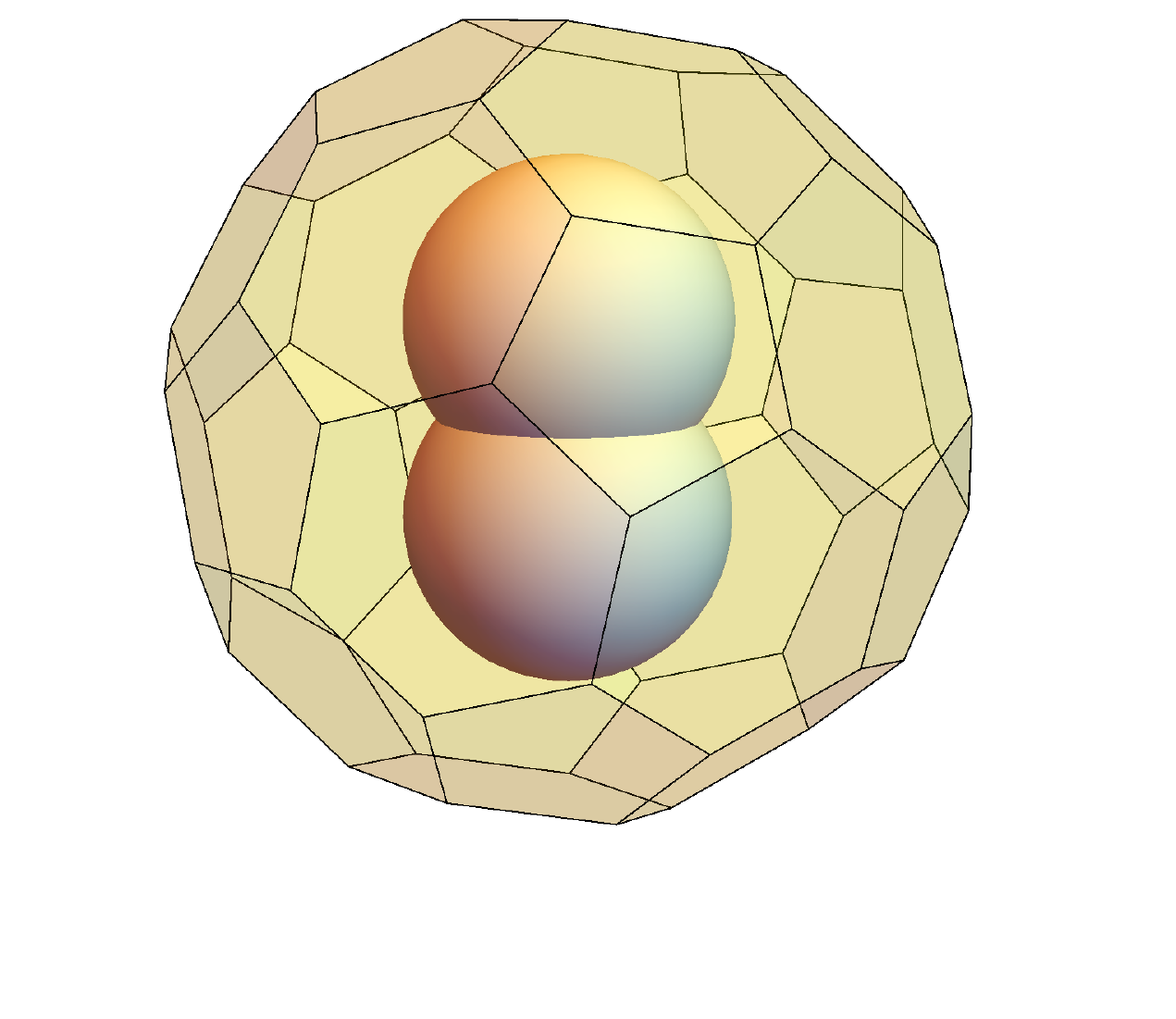}
\end{center}
\caption{Pictorial representation of a H$_2$ molecule trapped inside a fullerene (electron in carbon atoms not shown)\label{h2c60}}
\end{figure}

As an example of application to the realm of molecular physics, we will discuss a new algebraic theory of endohedrally confined diatomic molecules. 
The buckminsterfullerene molecule, or simply fullerene C$_{60}$, is composed of 60 carbon atoms placed
at the vertices of a icosahedral solid. The interior is almost hollow, except for the tails of the electron wavefunctions, and large enough to host other small chemical species, such as atoms and diatomic molecules.
These are said to be endohedrally confined and the whole system is called a supramolecular complex, indicated with the @ symbol.
For example the notation H$_2$@C$_{60}$ refers to a diatomic hydrogen molecule trapped inside a fullerene as in Fig. \ref{h2c60}. These systems are not bound by electric charge as in  ionic molecules, nor by the mutual exchange of electrons as in covalent molecule, nor by some effective dipole-dipole (or higher order) potential as in Van der Waals molecules. They are simply trapped by spatial confinement and the dimension of the spherical cage (the icosahedral symmetry can be disregarded to all practical effects, unless one is looking for tiny splittings due to the highly symmetrical environment) is small enough that the motion of the trapped molecule inside it exhibits quantization and this quantization quite significantly alters the spectral properties.
These systems have been produced thanks to an ingenious series of chemical reactions, known as 'molecular surgery' \cite{Koma}, that open a hole in the fullerene, force the gas in by applying pressure and then seal the molecule back to its original shape. Thus endohedrally confined species can be synthesized in the range of milligrams, well enough for experiments.

Infrared absorption, nuclear magnetic resonance and inelastic neutron scattering experiments have been performed and the status of both spectroscopic knowledge and traditional theoretical models is quite advanced (See \cite{Mamo, Horse, h2} and references therein). We have seen the opportunity to simplify and generalize the theoretical approaches to this system with an algebraic theory, inspired to common treatments in the nuclear physics domain.  The theoretical treatment  starts from the description of the free diatomic molecule as in the Vibron model of Iachello and Levine\cite{Iach}. The rotations and vibrations of such a molecule can very well be described with a $u(4)$ Lie algebra arising from a scalar ($s$, $L=0$) and a vector ($p$, $L=1$) bosons. The dynamical symmetries originating in the unitary algebra $u(4)$ and ending in the orthogonal algebra of the angular momentum $so(3)$, describe a large class of possible hamiltonians ranging from rigid structures with prevalent vibrational motion to easily deformable floppy molecules. The free hydrogen molecule is well described in the so-called $so(4)$ limit of the vibron model \cite{Piet}.
We have coupled this well-known analytic description to another Lie algebra, $u(3)$ that is used to model the effect of spatial confinement inside the cage. The three components of the vector boson $q$ (with $L=1$) are related to the displacement of the molecule as a whole from the equilibrium point at the center of the cage.
 
The dynamical symmetry that we consider thus amounts to  
\begin{equation}
\begin{array}{cccccc}
 u_p(4) \oplus u_q(3) & \supset  &so_p(4) \oplus u_q(3)  \supset& so_p(3) \oplus so_q(3)  & \supset so_{pq}(3)& \supset so_{pq}(2)   \\
 ~N_p    \hfill N_q~  &          &~\omega\hfill ~              &   ~J     \hfill L~      &       \Lambda    &         M_\Lambda  
\end{array}~,
\label{basis}
\end{equation}
where the quantum numbers associated with the Casimir
operators of each algebra are also given. As usual, $\omega$ is
replaced by the vibrational quantum number $v$ through $v={\textstyle
  \frac{1}{2}}(N_p-\omega)$. 
The set of quantum numbers $(vJN_qL\Lambda)$ can be used to label the states of the system.

The total Hamiltonian is
\begin{equation} 
\hat H_{endo} = \hat H_{u_p(4)} + \hat H_{u_q(3)} + \hat H_{Coupl}~,
\label{totH}
\end{equation}
where the first term is the vibron model Hamiltonian for
rotations and vibrations of a diatomic molecule, the second is the quantized motion of the molecular
center-of-mass inside the cage, and the last term includes molecule-cage couplings.

The $u(4)$ vibron model Hamiltonian can be modeled as follows
\begin{equation}
\hat H_{u_p(4)}= \hat H_{so(4)} + \hat H_{Dun}~,
\label{hup4}
\end{equation}
in terms of a $so(4)$ dynamical symmetry plus higher-order Dunham-like corrections:
\begin{eqnarray}
\hat H_{so(4)} &= E_0 +\beta\, \hat C_2[so_p(4)] + \gamma\, \hat C_2[so_p(3)]~,\label{so4ham}\\
\hat H_{Dun} &= \gamma_2\hat C_2[so_p(3)]^2+\kappa\,  \hat C_2[so_p(4)]\hat C_2[so_p(3)]~.\label{corrham}
\end{eqnarray}
giving the energy formula
\begin{equation}
E_{u_p(4)}= E_0 +\beta\, \omega (\omega+2) + \gamma\, J(J+1) + \gamma_2\Bigl[ J(J+1)\Bigr]^2+\kappa\,  \Bigl[ \omega(\omega+2)J(J+1)\Bigr]~,
\label{enII}
\end{equation}
where $\omega= N_p, N_p-2,\ldots, 1 ~{\mbox or }~  0$ and  $J=0,1,\ldots, \omega $.

The $u_q(3)$ dynamical symmetry gives instead
\begin{equation}
  \hat H_{u_q(3)}= a\, \hat C_1[u_q(3)] +b\, \hat C_2[u_q(3)]+c\, \hat C_2[so_q(3)]~,
  \label{cageham}
\end{equation}
that can be immediately translated into eigenenergies, using the eigenvalues of Casimir operators:
\begin{equation}
  E_{u_q(3)}= a\, N_q+  b\, N_q^2+ c\, L(L+1)~,
  \label{hospec}
\end{equation}
where $N_q$ is the number of quanta and $L$ is the orbital angular momentum.

Finally the cage-molecule couplings could be of several types, but we have selected those that might be relevant and proposed the following interaction:
$$
\hat H_{Coupl} = \vartheta_{pq} [\hat Q_p^{(2)}\times \hat Q_q^{(2)}]^{(0)} + \vartheta_{pqw} \left[\hat C_2[so_p(4)][\hat Q_p^{(2)}\times \hat Q_q^{(2)}]^{(0)}+[\hat Q_p^{(2)}\times \hat Q_q^{(2)}]^{(0)}\hat C_2[so_p(4)]\right]+$$
\begin{equation}
+ v_{pq} \hat C_1[u_q(3)]\hat C_2[so_p(4)]~,\label{hcoup}
\end{equation}
where the parameters can
be adjusted to model the effects of the coupling. The  $[\hat Q_p^{(2)}\times \hat Q_q^{(2)}]^{(0)}$
quadrupole-quadrupole interaction lifts the degeneracy of
$\Lambda \ne 0$ multiplets, giving the correct and somewhat unusual ordering seen in experiments.
The matrix elements can be found in Ref. \cite{h2}. Formulas (\ref{enII}) and (\ref{hospec}) plus the diagonalization of the coupling terms (\ref{hcoup}) give a practical and accurate way of modeling the caged H$_2$ molecule.

\begin{figure}
\begin{center}
\includegraphics[width=0.9\textwidth, bb=21 47 550 420]{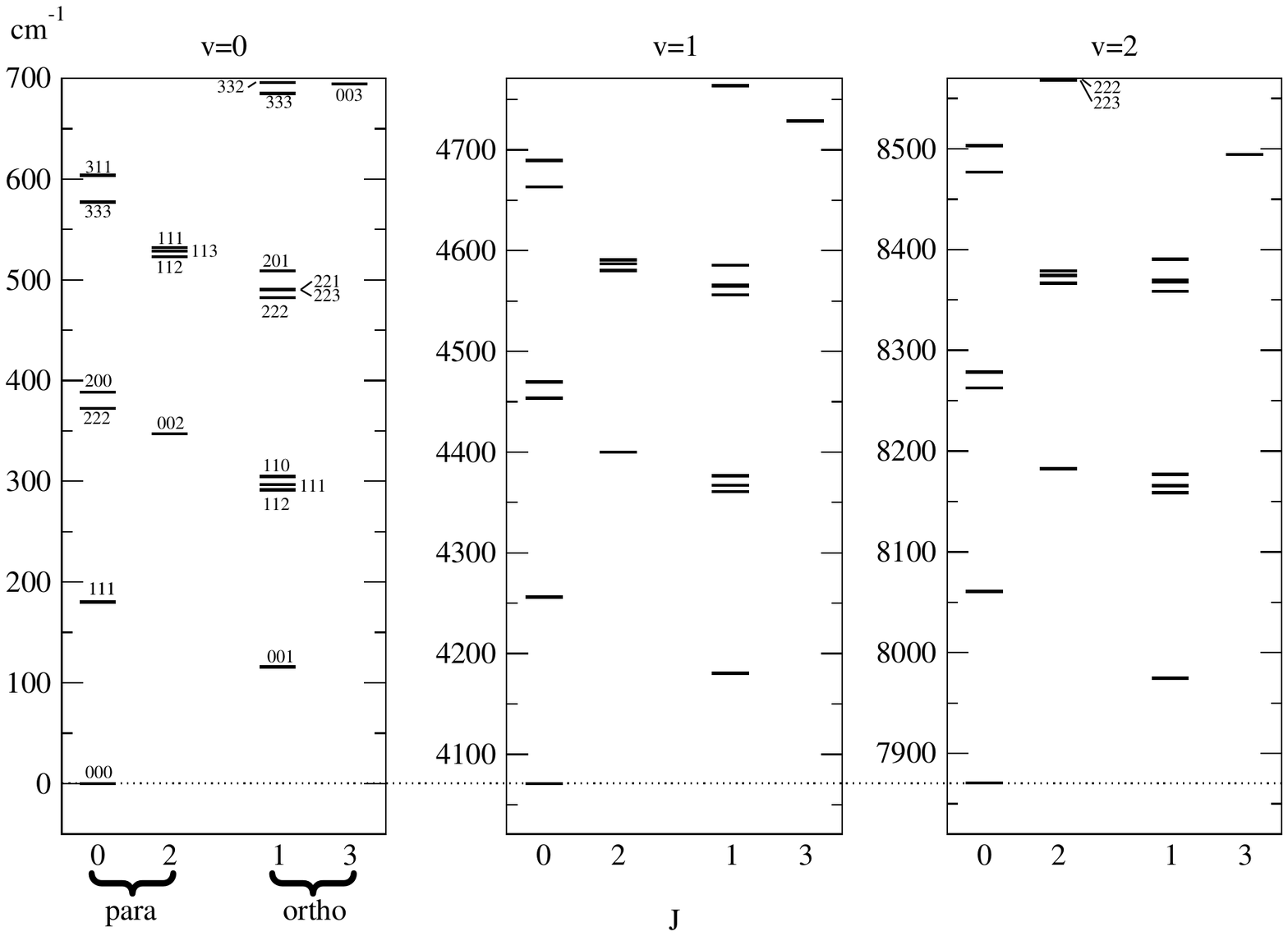}
\end{center}
\caption{Theoretical rotational-vibrational spectrum of H$_2@$C$_{60}$. 
The three panels show the lowest states built on top of the vibrational excitations with $v=0,1$,
  and $2$. States are divided into \textit{para} and
  \textit{ortho} states and labeled by $J$ on the
  horizontal axis and $N_qL\Lambda$ on each state. \label{spect}}
\end{figure}

Extensive numerical investigations and minimization of theory-to-experiments least square differences, using a body of about 70 transition lines, has lead us to a set of parameters that improves the overall theoretical description with respect to competing theories (for example we have only 10 parameters for all possible vibrational quantum number $v$ bands, while others have been fitting these bands separately with a larger number of parameters) and allows to suggest that a few published level assignments were probably incorrect. But the greatest advantage is that it gives a unique theoretical framework that contains all symmetry-allowed terms, even without the precise knowledge of each atom-atom interaction. 
The lowest portion of the spectrum, that differs in several respects from that of the free molecules (energy shifts and splitting of energy levels) is given in Fig. \ref{spect}, where one can see several effects, for example the lowering of certain groups of states with increasing value of $v$ or the increasing splitting with increasing values of $L$, etc.

Our theory is at the same time easier and more universal than previous fits based on direct solution of the Schroedinger equation and allows to make predictions for yet-to-be-measured states. It might seem that this type of theoretical models based on Lie algebras do not contain directly an expression for the interactions between the constituents, i.e. some $V_{ij}(r_{ij})$, and this is disturbing to many, but the advantage is that this information is not needed at all: if the hamiltonian complies with a certain symmetry (dynamical symmetries from higher/bigger algebra down to angular momentum SU(3) or SU(2)) the spectrum is automatically analytically known and the phenomenology boils down to the coefficients of the energy formulas. This is a very desirable tool.

\section{Algebraic model for hypernuclei}
We turn our attention now to applications in the realm of particle physics by describing an algebraic theory that applies to hyperons bound in a nuclear environment.

A series of experiments at KEK-PS SuperKaonSpectrometer measured the yield of $\Lambda$ particles inside medium-mass or heavy nuclei. These particles have been produced in different angular momentum states inside the nucleus, but a complete theoretical understanding is still incomplete \cite{Hotchi}.
In particular, single-particle $\Lambda$ states with orbital angular 
momentum ranging from $l=0$ to $l=3$ have been 
clearly identified in medium-heavy $^{89}_\Lambda$Y hypernucleus using the 
$(\pi^+,K^+)$ reaction spectroscopy \cite{Hashi,Hotchi}. 
These single-particle levels have been theoretically analyzed with the 
distorted-wave impulse approximation (DWIA) based on shell 
model calculations \cite{Motoba08}.

There have been nice theoretical works based on extensions of the shell-model that include that $\Lambda$ degree of freedom and we can propose a model based on Lie algebras that is, in general, equivalent to an anharmonic oscillator. In this approach, single-particles levels are 
classified according to the underlying symmetries. We use a simple $su(3)$ bosonic symmetry for the anharmonic oscillator and an even simpler $su(2)$ fermionic symmetry for the inclusion of the hyperonic degree of freedom.
The energy of each level is then given in terms of 
expectation values of Casimir operators associated with the following 
dynamical symmetry 
\be
u(3/2) \supset u_B(3)\times u_F(2) \supset so_B(3)\times su_F(2)
\ee
that is the only possible dynamical symmetry arising in this simple scheme.

We consider the following Hamiltonian:
\be
\hat H_{\rm Hyp} = \hat H_{\rm u(3)} + \hat H_{\rm u(2)} + \hat V_{\rm int},
\label{Hhyp}
\ee
where 
\be
\hat H_{\rm u(3)}= \alpha\, \hat C_1(u(3)) +\beta\, \hat C_2(u(3))
+ \gamma\, \hat C_2(so_L(3))
\ee
represents the anharmonic oscillator 
with $\alpha, \beta$, and $\gamma$ as free parameters. 
The Casimir operators are given by 
$\hat C_1(u(3))=\hat N$, $\hat C_2(u(3))=\hat {N^2}$ and 
$C_2(so_L(3))=\hat {\vec{L}^2}$. 
The spectrum is:
\be
E_{\rm u(3)}=  \alpha N+  \beta N(N+2)+ \gamma L(L+1)
\label{hospec2}
\ee
where $N$ is the number of quanta and $L$ is the orbital 
angular momentum of the $\Lambda$ particle. 

The fermionic part of the Hamiltonian is: 
\be
\hat H_{\rm u(2)} =A \hat C_1\bigl(u(2)\bigr) + B\hat C_2\bigl(u(2)\bigr),   
\ee
and the energy spectrum is 
\be
E_{\rm u(2)} = A \langle C_1\rangle  + B\langle C_2\rangle. 
\ee
The representations of $u(2)$ are given by a pair of numbers $[\lambda_1, \lambda_2]$. 
Since there are only three possible fermionic states, 
the formula can take the values: 
\be
E_{0} =0, \qquad
E_{\Lambda} = A+2B, \qquad
E_{\Lambda\Lambda} = 2A+2B.  
\label{syst}
\ee
The final energies for hypernuclei are thus
\be
E_{\rm Hyp} = 
\alpha N+  \beta N(N+2)+ \gamma L(L+1) +E_{n\Lambda},
\label{Ehyp}
\ee
where the last term is given by Eq. (\ref{syst}), depending on 
the number of $\Lambda$ particles in the system. 

An application of our formula is made on the $^{89}_\Lambda$Y 
and $^{51}_\Lambda$V hypernuclei. 
The measured cross-sections (integrated in the 2$^o$-14$^o$ range) 
for the ($\pi^+, K^+$) reaction leading to the formation 
of the hypernuclei 
show several peaks as a function of energy \cite{Hotchi}, 
that are interpreted 
as corresponding to different 
angular momentum states of the $\Lambda$ particle. 

\begin{figure}[!t]
\includegraphics[width=0.8\textwidth, clip=, bb= 00 00 700 600]{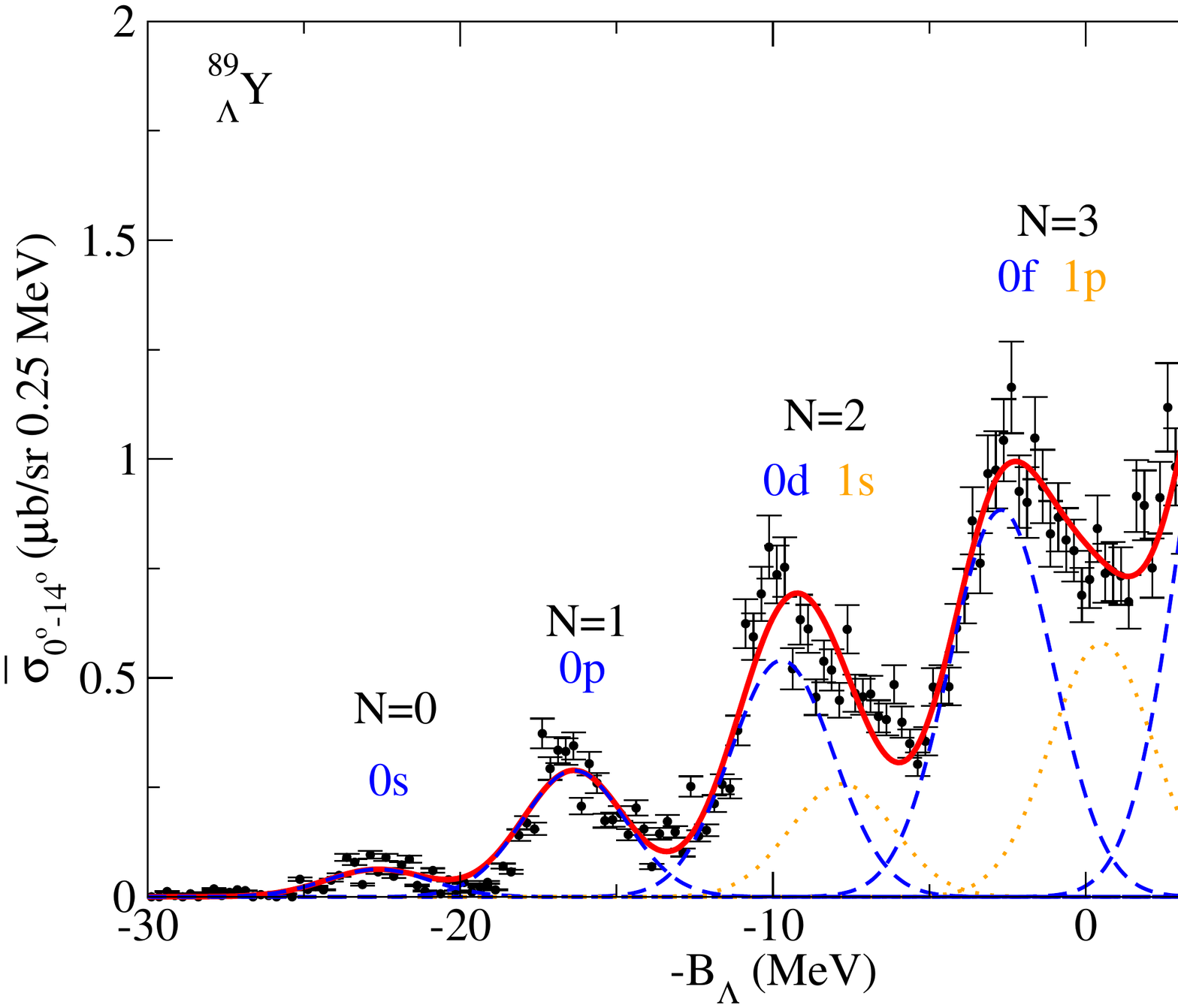}
\caption{
Experimental hypernuclear mass spectra of $^{89}_\Lambda$Y  and fit
based on the algebraic model. } 
\label{fit0}
\end{figure}

\begin{figure}[!t]
\includegraphics[width=0.8\textwidth, clip=, bb= 00 00 600 500]{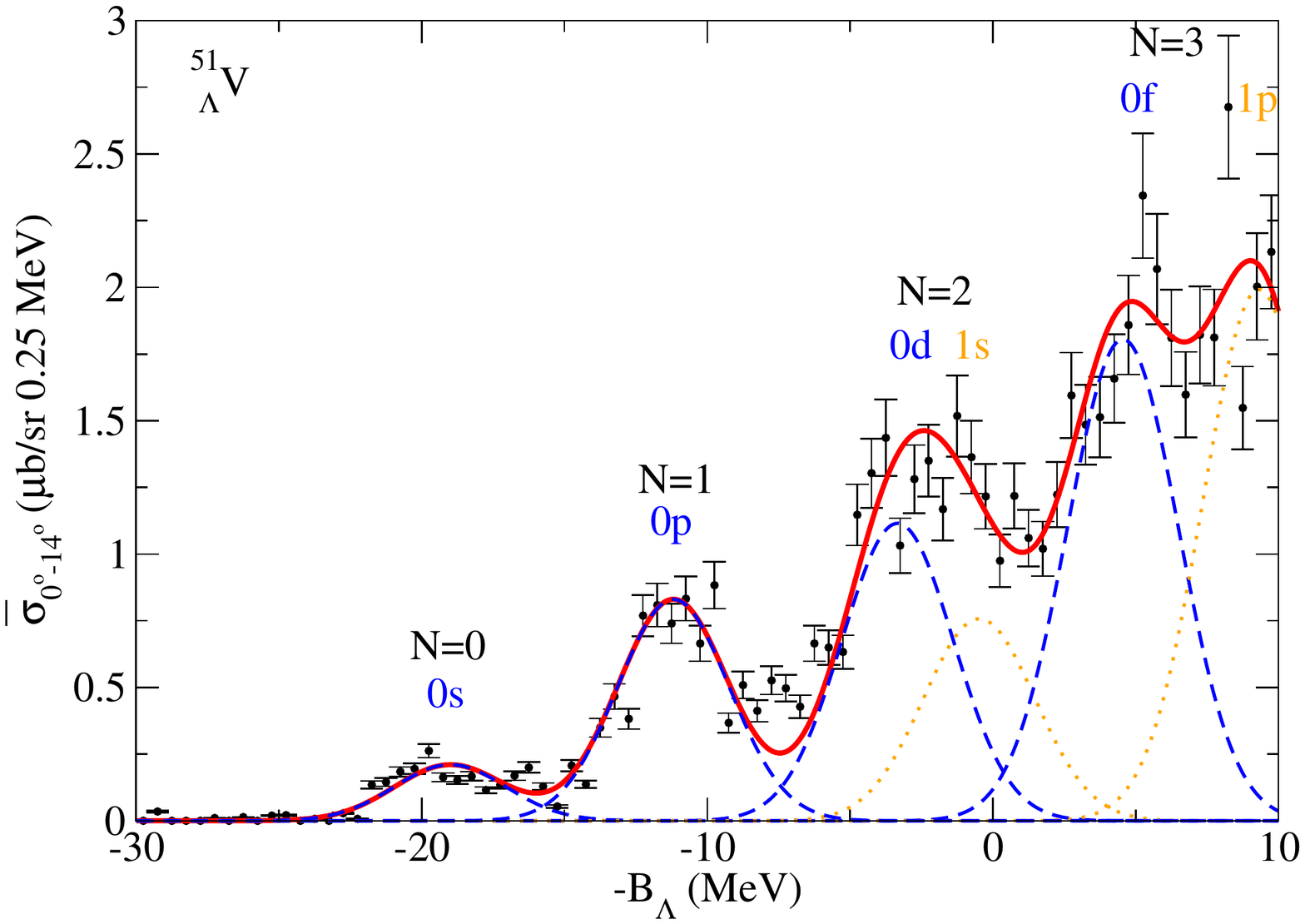}
\caption{Same as Fig. \ref{fit0}, but for the $^{51}_\Lambda$V hypernucleus.}
\label{V-fit0}
\end{figure}

Our fit has the same main ingredients of the work of Hotchi, namely gaussian functions with parametric width, but the advantage is that the centroids are related to our energy formula (details in \cite{hyp}). In general we have fits of the same statistical value, but with a smaller number of parameters and with a theoretical framework for the peaks. Our results are summarized in Fig. \ref{fit0} and \ref{V-fit0} and Tables \ref{par89Y} and \ref{par51V}, where one can read off the best parameters to fit the data within the algebraic model.

\begin{table}[!t]
\begin{center}
\begin{tabular}{cc|cc|cc|cc}
\hline
\hline
$a_{00}$ & 1.03458 & $a_{11}$ & 4.774  & $a_{20}$ & 4.26848 & $a_{22}$ & 8.93652\\
$a_{31}$ & 9.57975 & $a_{33}$ & 14.6199 & $a_{42}$ & 21.4563 & $a_{44}$ & 22.7715\\
\hline
$\alpha$ & 5.39547 & $\beta$  &0.506972 & $\gamma$ & $-$0.321663 
& $E_\Lambda$ & $-$22.6373 \\
\hline
\hline
\end{tabular} 
\caption{Parameters for the best fit of the 
empirical mass spectra of $^{89}_\Lambda$Y. 
The parameters $\alpha, \beta, \gamma$ and $E_\Lambda$ are in MeV, while $a_{NL}$ are in $\mu$b MeV.}
\label{par89Y}
\end{center}
\end{table}

One can see that our algebraic model nicely reproduces the data and gives an interpretation of peaks according to (an)harmonic oscillator quantum numbers, describing the major shells and the angular momentum energy shifts. One would ideally include higher order effects (spin-orbit coupling for example) with the aim of getting even better agreement with data, but the present level of accuracy seems adequate to the energy resolution that the data points display.
Of course this model is very simple and it might become more and more incorrect as long as one goes close to the $\Lambda$ separation threshold (inasmuch as the harmonic oscillator shell model has been abandoned in favour of the Wood-Saxon potential well), but it gives a nice way 
of classifying states according to a simple theoretical scheme and it can easily be extended to more species or to particles with higher spin.

\begin{table}[!t]
\begin{center}
\begin{tabular}{cc|cc|cc|cc}
\hline
\hline
$a_{00}$ & 4.11173 & $a_{11}$ & 16.2407  & $a_{20}$ & 14.8101 & $a_{22}$ & 21.8096\\
$a_{31}$ & 38.9592 & $a_{33}$ & 35.3016 &  & \\
\hline
$\alpha$ & 7.2807 & $\beta$  &0.495241 & $\gamma$ & $-$0.476428 
& $E_\Lambda$ & $-$19.003 \\
\hline
\hline
\end{tabular} 
\caption{Same as Table. \ref{par89Y}, but for the $^{51}_\Lambda$V hypernucleus.}
\label{par51V}
\end{center}
\end{table}

\section{Conclusions}
We have shown applications of nuclear physics theories or models inspired to typical nuclear physics model in very different branches of physics like molecular physics, nuclear astrophysics and hypernuclear physics, with energies ranging from meV to keV to GeV. Our project, aimed at interdisciplinary applications of nuclear theories to other fields, has generated a large amount of new (and hopefully interesting) physics and we are determined to continue on this track.
Our future endeavours will be the proposal of a new method to solve the non-relativistic Schroedinger equation, with the aim of applying it to atomic. molecular and nuclear physics cases.

\ack
All the collaborators and coauthors of the researches summarized in this paper are thankfully acknowledged. {\it In:Theory}, PRAT Project n. CPDA154713, Univ. Padova (Italy).

\end{document}